\documentclass[a4paper,fleqn,usenatbib]{mnras}

\usepackage{newtxtext,newtxmath}

\usepackage[T1]{fontenc}
\usepackage{ae,aecompl}

\usepackage{graphicx}	
\usepackage{amsmath}	
\usepackage{amssymb}	
\usepackage{pdflscape}

\title[Confirmation of water fountains]{Interferometric confirmation of ``water fountain'' candidates}

\author[G\'omez et al.]{
Jos\'e F. G\'omez,$^{1}$\thanks{On sabbatical leave at Laboratoire Lagrange, Universit\'e de Nice Sophia-Antipolis, CNRS, Observatoire de la C\^ote d'Azur, F-06304 Nice, France}\thanks{E-mail: jfg@iaa.es}
Olga Su\'arez,$^{2}$
J. Ricardo Rizzo,$^{3}$
Lucero Uscanga,$^{4}$
\newauthor
Andrew Walsh,$^{5}$
Luis F. Miranda$^{1}$
and Philippe Bendjoya$^{2}$
\\
$^{1}$Instituto de Astrof\'{\i}sica de Andaluc\'{\i}a, CSIC, Glorieta de la Astronom\'{\i}a s/n, E-18008 Granada, Spain\\
$^{2}$Laboratoire Lagrange, UMR 7293, Universit\'e de Nice Sophia-Antipolis, CNRS, Observatoire de la C\^ote d'Azur, F-06304 Nice, France\\
$^{3}$Centro de Astrobiolog\'{\i}a (INTA-CSIC), Ctra. M-108, km. 4, E-28850 Torrej\'on de Ardoz, Spain\\
$^{4}$Departamento de Astronom\'{\i}a, Universidad de Guanajuato, A.P. 144, 36000 Guanajuato, Gto., Mexico\\
$^{5}$International Centre for Radio Astronomy Research, Curtin University, GPO Box U1987, Perth WA 6845, Australia
}

\date{Accepted XXX. Received YYY; in original form ZZZ}

\pubyear{2016}

\begin{document}
\label{firstpage}
\pagerange{\pageref{firstpage}--\pageref{lastpage}}
\maketitle

\begin{abstract}
Water fountain { stars} (WFs) are evolved { objects} with water masers tracing high-velocity jets (up to several hundreds of km s$^{-1}$). They could represent one of the first manifestations of collimated mass-loss in evolved objects and thus, be { a key} to understanding the shaping { mechanisms} of planetary nebulae. Only 13 objects { had been confirmed so far} as WFs with interferometer observations. We present new observations with the Australia Telescope Compact Array and archival observations with the Very Large Array of four objects that are considered to be WF candidates, mainly based on single-dish observations. We confirm IRAS 17291-2147 and IRAS 18596+0315 (OH 37.1-0.8) as bona fide members of the WF class, with high-velocity water maser emission {  consistent with} tracing bipolar jets. We argue that IRAS 15544-5332 has been wrongly considered { as} a WF in previous works, since we see no evidence in our data nor in the literature that this object harbours high-velocity water maser emission. In the case of IRAS 19067+0811, we did not detect any water maser emission, so its confirmation as a WF is still pending. With the result of this work, there are 15 objects that can be considered confirmed WFs. We speculate that there is no significant physical difference between WFs and obscured post-AGB stars in general. The absence of high-velocity water maser emission in some obscured post-AGB stars  could be { attributed to} a variability or orientation effect.
\end{abstract}




\section{Introduction}

``Water fountain'' { stars (WFs)} are evolved objects that show { high-velocity water masers, with at least some maser components tracing collimated jets} \citep[see][for reviews]{ima07,des12}. 
The velocity spread in
their water maser spectra is typically $\ge 80$ km s$^{-1}$, and can
be as large as $\simeq 500$ km s$^{-1}$ \citep{gom11}. 
{ These spectra indicate the presence of motions with velocities much higher than those of the expanding circumstellar envelopes of objects in the Asymptotic Giant Branch
(AGB), which are estimated to be $\simeq 5-20$ km s$^{-1}$, based on the velocity separation between the the components in double-peaked OH maser spectra \citep[e.g.,][]{tel89,sev97}.}
WFs are
in the short transition phase between the { AGB} and the planetary nebula (PN) stage of low- to intermediate-mass stars ($< 8$ M$_\odot$). Their being in this particular
evolutionary stage, together with the short dynamical { timescales} of the
maser jets  \citep[5-100 yr;][]{ima07}, and their relatively strong optical
obscuration \citep*{sua08} indicate that WFs may represent one of the first
manifestations of collimated mass-loss in evolved stars.
Therefore, they are key objects to understand when and how the
spherical symmetry seen in AGB stars and earlier stages is broken to
give rise to the spectacular variety of morphologies displayed in the
PN phase.

Only around 15 sources have been identified so far as possible WFs \citep{des12}. Of these, 13 have been confirmed with interferometric
observations. The rest have been identified { as candidate WFs} with single-dish spectra. A usual criterion to identify the { possible} presence of  high-velocity gas in these cases is the { detection} of water maser emission outside the velocity range of OH { \citep[e.g.,][]{yun13}}, considered to trace the expansion of the circumstellar envelope.  This criterion is useful even in the cases where the jet traced by the water maser emission moves close { to} the plane of the sky and thus, the measured spread of radial velocities of water masers are modest ($\la 50$ km s$^{-1}$). However,
given the relatively low angular resolution of single-dish observations, it
is impossible to ascertain whether the multiple water maser peaks, spread over
large velocities, actually arise from { the same} source or they are the
superposition of emission from close, but different sources. { Interferometric observations, with a high positional accuracy, are necessary to ascertain that all maser components are associated with the same evolved object.}

The confirmation of new objects as WFs is important, given the scarcity of members of this class. This small number of members may indicate either that the WF phase is short, and/or that only a small fraction of stars become WFs.
Increasing the database of this type of objects may help us to understand unsolved questions, such as whether most low and intermediate-mass stars go through a WF phase. Moreover, monitoring of the identified objects would give us clues as to whether WF characteristics occur during a very limited time in stellar evolution, or whether they are recurrent episodes along a longer period of time. 

In this paper we present interferometric water maser observations toward four sources that are considered to be WF candidates, given the presence of high-velocity water maser emission in single-dish spectra, but for which no interferometric confirmation was yet available.

\section{Observations}

Our observations were carried out on 2014 September 28 with the Australia Telescope Compact Array (ATCA) in its H214 configuration, under the observational project C2954. We used the Compact Array Broadband Backend (CABB) correlator \citep{wil11} in its CFB 1M-0.5k mode, which provides, in each of the two available intermediate frequency bands (IFs), a broad bandwidth of 2 GHz sampled over 2048 channels of 1-MHz width, plus up to 16 spectral zooms to study spectral lines with a resolution of 0.5 kHz. In the case of our L/S-band observations, we centered the 2 GHz broad band of CABB at 2100 MHz for both IFs (thus providing redundant data), and twelve of the
{ broadband} channels (four groups of 3 channels) were zoomed
in to observe
spectral lines. With this setup, we observed the OH ground-level
transitions of rest frequencies 1612.2309, 1665.4018,
1667.3590, and 1720.5299 MHz, with { an effective frequency range} of
2 MHz each, sampled over 4096 channels of 0.5 kHz (total
velocity coverage and resolution of 381 and 0.09 km s$^{-1}$ at
1612 MHz, respectively). For our K-band observations, the two IFs were centered at 20000 and 22000 MHz. We combined 16 zoomed-in broadband  channels to observe the $6_{16}-5_{25}$ transition of the H$_2$O molecule (rest frequency 22235.08 MHz), providing a total 
velocity coverage and resolution of 114 and $6.6\times 10^{-3}$ km s$^{-1}$, respectively. These H$_2$O-maser data were Hanning-smoothed, up to a velocity resolution of 0.066 km s$^{-1}$. Source  PKS 1934-683 was used to set the absolute flux scale at all frequencies, whereas the spectral response of the instrument was calibrated with sources 	PKS 1934-638 and 3C 279 at L/S and K bands, respectively. A summary of other observational parameters is given in Table \ref{tab:obs}. { Note that the source IRAS 19067+0811 was not observed at L/S band}. The velocities in this paper are given with respect to the kinematic definition of the Local Standard of Rest (LSRK).

\begin{table*}
	\centering
	\caption{{ ATCA} observational and imaging parameters.}
	\label{tab:obs}
	\begin{tabular}{lllllccl} %
		\hline
		Target & \multicolumn{2}{c}{Phase centre} &   Frequency & Gain calibrator &  $\theta_{\rm fwhm}$$^a$ & $\theta_{\rm pa}$$^b$  & $\sigma_{\rm lf}$$^c$\\
			  			&	R.A.(J2000)	& 	Dec(J2000)	& (MHz)	&				& (arcsec$^2$) & (deg) & (mJy)\\
		\hline
		IRAS 15544$-$5332	& 15:58:18.40 	& $-53$:40:40.0 & 1612 & PMN J1617-5848 & $157\times 100$ & +80 & 16 \\
							&				&				& 1665 & \multicolumn{1}{c}{''} & $146\times 97$& +77 & 13\\
							&				&				& 1667 & \multicolumn{1}{c}{''} & $146\times 97$& +79 & 12\\
							&				&				& 1720 & \multicolumn{1}{c}{''} & $139\times 91$& +76 & 10\\
						& 				& 				& 22235 & PMN J1534-5351 & $11.4\times 7.6$ & $-89$ & 6 \\
		IRAS 17291$-$2147	& 17:32:10.10	& $-21$:49:59.0 & 1612 & PMN J1713-2658 & $134\times 98$ & $-89$ & 7\\
							&				&				& 1665 & \multicolumn{1}{c}{''} & $135\times 93$ & $-89$ & 7\\
							&				&				& 1667 & \multicolumn{1}{c}{''} & $135\times 93$ & $-89$ & 9\\
							&				&				& 1720 & \multicolumn{1}{c}{''} & $128\times 94$ & $+90$ & 8\\
						&				&				& 22235 & PMN J1709-1728 & $10.3\times 7.8$& +71 & 4\\
		IRAS 18596+0315 & 19:02:06.28	& +03:20:16.3 	& 1612 & JVAS J1907+0127& ---$^d$ \\
						&				&				& 1665 & \multicolumn{1}{c}{''} & ---$^d$ \\
						&				&				& 1667 & \multicolumn{1}{c}{''} & ---$^d$ \\
						&				&				& 1720 & \multicolumn{1}{c}{''} & ---$^d$ \\
						&				&				& 22235 & QSO J1851+005  & $12.9\times 8.0$& +50 & 5\\
		IRAS 19067+0811 & 19:09:07.47	& +08:16:22.5	& 22235 & VCS2 J1856+0610& $11.2\times 7.7$& +40 & 11\\
		\hline
	\end{tabular}
	
	$^a$ Full width at half maximum of the synthesized beam.\\
	$^b$ Position angle (north to east) of the synthesized beam.\\
	$^c$ {  Rms noise in line-free channels. The noise in channels with strong maser emission is higher, due { to} limitations in dynamic range.}\\
	$^d$ {  OH observations in IRAS 18596+0315 did not have enough uv coverage to obtain images.}
\end{table*}

Data calibration and initial imaging was carried out with the {\sc miriad} package. { For the broad-band continuum data, we obtained images in each of the two observed frequency ranges, using multifrequency synthesis over the whole bandwidth (2 GHz centred at 2.1 GHz and 4 GHz centred at 21 GHz), and natural weighting of visibilities.  
In the case of the maser lines,} once the spectral channel with maximum emission for each line and source was identified, self-calibration was attempted on that channel, using the {\sc aips} package. When this self-calibration was successful, the computed phase and amplitude corrections { were} then applied to all channels. Final images were obtained with {\sc aips}, using a Briggs weighting with robust parameter equal to zero. All baselines with antenna 6 (located at 4 km from the array centre) were excluded from the self calibration and imaging, { since the uv-coverage provided by these long baselines was not enough for mapping}. 

We also downloaded all H$_2$O-maser data observed with the Karl G. Jansky Very Large Array (VLA) of the National Radio Astronomy Observatory (NRAO) toward the source IRAS 18596+0315 (projects AG330, AR277, AG461, AG573). Calibration and further data processing were carried out with AIPS. Images were obtained with a robust parameter equal to zero.

Maser components are defined as distinct peaks in the spectra. For each component we measured their spatial location by fitting a 2-dimensional Gaussian to the emission in the channel with maximum emission for this component, {  using task {\sc jmfit} of {\sc aips}. All maser components are unresolved at the angular resolution of our observations. For unresolved sources, the relative (1 $\sigma$) positional accuracy for the components in each data set is $\sigma_{\rm pos} \simeq \theta_{\rm fwhm}/(2 {\rm snr})$ \citep{con98}, where $\theta_{\rm fwhm}$ is the full width at half maximum (fwhm) of the synthesized beam}, and ${\rm snr}$ is the signal-to-noise ratio of the { emission}. Thus, it is possible to determine morphological distributions at scales well below 1 arcsec in the case of water masers. The positional errors given in the tables in this paper reflect these relative positional accuracies, {  and were calculated with task {\sc jmfit}, which implements precise error estimates following \cite{con97}.} The errors in absolute astrometry are generally larger (typically on the order of 10\% of the synthesized beam). { A comparison with infrared positions (see sec. \ref{sec:17}) suggests that our absolute astrometry could be in error by $\simeq 3$ arcsec.}

All uncertainties in this paper are given at a 2$\sigma$ level. The quoted uncertainties in flux densities reflect the errors due to noise in the maps, not any possible error in the determination of the absolute flux density scale using PKS 1934-638. { Flux density} upper limits are given at a 3$\sigma$ level.

\section{Results}

\label{sec:results}
Water maser emission (Table \ref{tab:h2o}) was detected in IRAS 15544-5332 and IRAS 17291-2147. OH maser emission (Table \ref{tab:oh}) was detected in  IRAS 18596+0315 only (note that we did not make any OH observations toward IRAS 19067+0811), both at 1612 and 1667 MHz.  A detailed description of the results on each source is given in the following subsections. In addition to our target sources, OH maser emission at 1612, 1665, and 1667 MHz was detected from other sources within the observed fields (Appendix \ref{sec:app}).

{ We did not detect any continuum emission in the broadband data. Three sigma upper limits for the 4-GHz  bandwidth centred at 23 GHz were 40 $\mu$Jy for IRAS 15544-5332 and 30 $\mu$Jy for the other three sources. For the 2-GHz bandwidth centred at 2.1 GHz, we obtained $3\sigma$ upper limits of 12 mJy for IRAS 15544-5332 and IRAS 18596+0315 { (this noise level was mainly due to confusion by galactic background emission)}. No map at 2.1 GHz was obtained for IRAS 17291-2147, since most of its data had to be flagged out due to the presence of radio frequency interference. }

\begin{table*}
	\centering
	\caption{Parameters of water maser emission.}
	\label{tab:h2o}
	\begin{tabular}{lllllllll} %
		\hline
Source & $S_\nu$$^a$ & $V_{\rm peak}$$^b$  & R.A.(J2000)$^c$ & Dec(J2000)$^c$ & Error$^d$ & $\int S_\nu dV$$^e$ & $V_{\rm min}$$^f$ & $V_{\rm max}$$^g$ \\
		& (Jy) &	(km s$^{-1}$)  & 					&			& (mas) &(Jy km s$^{-1}$) & (km s$^{-1}$) & (km s$^{-1}$) \\
		\hline
IRAS 15544-5332 & $0.183\pm 0.013$ & $-73.16$ & 15:58:18.89 & $-53$:40:38.6 & 300 & $0.092\pm 0.005$ & $-73.55$ & $-72.89$\\ 
IRAS 17291-2147 & $2.656\pm 0.011$ & $-16.77$ & 17:32:11.3749 & $-21$:50:02.846 & 23 & $6.72\pm 0.03$ & $-19.80$ & $+53.81$\\
IRAS 18596+0315 & $<0.016$\\
IRAS 19067+0811 & $<0.03$\\
		\hline
	\end{tabular}
	
	$^a$ Flux density of the maximum emission {  with its associated 2$\sigma$ uncertainty. Upper limits are given at a 3$\sigma$ level.}\\
	$^b$ LSR velocity of the maximum emission.\\
	$^c$ Position of the maximum emission.\\
	$^d$ 2$\sigma$ relative positional uncertainty between features in the same map. { Values of the coordinates are given up to this uncertainty. The uncertainty in absolute coordinates is estimated to be $\simeq 3$ arcsec (see text.)}\\
	$^e$ Integrated intensity of the source spectrum {  with its associated 2$\sigma$ uncertainty.}\\
	$^f$ Minimum velocity at which emission is present.\\
	$^g$ Maximum velocity at which emission is present.
\end{table*}

\begin{table*}
	\centering
	\caption{Parameters of OH maser emission.}
	\label{tab:oh}
	\begin{tabular}{lllllll} %
		\hline
Source & Frequency 	& $S_\nu$	& $V_{\rm peak}$	& $\int S_\nu dV$ & $V_{\rm min}$ & $V_{\rm max}$ \\
		& (MHz)		& (Jy) 		& (km s$^{-1}$) 	& (Jy km s$^{-1}$) & (km s$^{-1}$) & (km s$^{-1}$) \\
		\hline
IRAS 15544-5332 & 1612 & $<0.05$ \\
				& 1665 & $<0.04$ \\
				& 1667 & $<0.04$ \\
				& 1720 & $<0.03$\\
IRAS 17291-2147	& 1612 & $<0.022$\\
				& 1665 & $<0.022$\\
				& 1667 & $<0.03$\\
				& 1720 & $<0.023$\\
IRAS 18596+0315	& 1612 & $8.35\pm 0.06$ & $+76.8$ & $37.40\pm 0.07$ & $+73.5$ & $+103.3$  \\
				& 1665 & $<0.20$\\
				& 1667 & $1.29\pm 0.13$ & $+74.3$ & $2.34\pm 0.09$ & $+73.0$ & $+104.3$\\
				& 1720 & $<0.16$\\
		\hline
	\end{tabular}
\end{table*}

\subsection{IRAS 15544-5332 (OH 328.477-0.342)}

\label{sec:15}



IRAS 15544-5332 was first suggested as a possible WF by \cite{dea07}, based on single-dish observations. Their water maser spectrum  shows only a peak, at $\simeq-74$ km s$^{-1}$, which would normally not qualify as that of a WF, since these are usually identified by the large velocity spread of maser components. { However, the water maser component { is } $\simeq 40$ km s$^{-1}$ { away} from the velocity of the OH emission \citep*[$\simeq -118$ km s$^{-1}$;][]{sev97,dea04} at 1612 MHz.} \cite{dea07} assumed that the central OH velocity represent the stellar velocity, and the water maser component is part of the redshifted lobe of a WF jet. They expect its blueshifted counterpart at $-155$ km s$^{-1}$, but this was not detected.

ATCA observations by \cite{sev97} confirmed that the OH maser emission arises from IRAS 15544-5332 (OH maser position: R.A.(J2000) $= 15^h 58^m 18.830^ s$, Dec(J2000) = $-53^\circ 40' 40.20''$). 
However, there is no published interferometric observation of water masers, so there was no guarantee that the water maser emission arises from the same source. Moreover, there is no reported detection of the supposed blueshifted water maser components around $-155$ km s$^{-1}$.

\begin{figure*}
	\includegraphics[width=0.8\textwidth]{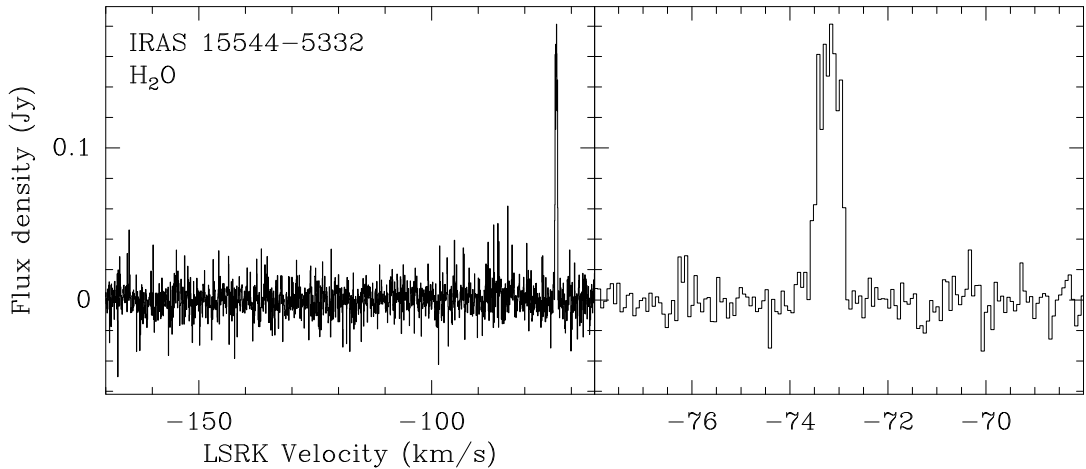}
	\caption{Water maser spectrum toward IRAS 15544-5332, shown with different velocity ranges.}
	\label{fig:spec_i15h2o}
\end{figure*}

In our observations, we detect a single peak of water maser emission at $-73.2$ km s$^{-1}$ (Table \ref{tab:h2o} and Fig. \ref{fig:spec_i15h2o}), but no OH emission (Table \ref{tab:oh}). { The derived OH upper limit at 1612 MHz (50 mJy) is significantly lower than the flux density of the emission reported by other authors at this frequency \citep[0.6-1.75 Jy]{sev97,dea04}, indicating a sharp decrease of the OH emission.} Our data confirm, for the first time, that the water maser emission arises from IRAS 15544-5332, since { the position we found} is { compatible with that reported for} the infrared { emission. The central source in different infrared catalogs is 2MASS J15581883-5340398  (nominal position 1.4 arcsec away from our H$_2$O maser position), WISEA J155818.84-534040.0 (1.4 arcsec away) and AKARI-IRC-V1 J1558187-534040 (2.4 arcsec away). The reported position of the OH maser emission \citep{sev97} is 1.7 arcsec away from our water maser position}. 


Note, however, that we did not detect any high-velocity water maser component. In fact, the assumption by \cite{dea07} that the velocity of OH emission (-118 km s$^{-1}$) corresponds to the stellar velocity is likely to be wrong. 
{ Interferometric observations of OH show a single peak at that velocity \citep{sev97}, and other spectral features seen with single-dish observations \citep{dea04} are probably the result of confusion from other sources, or diffuse interstellar emission.}
The OH spectra in AGB and post-AGB stars typically shows a double-peaked profile, tracing the approaching and receding part of the circumstellar envelope. Then, the stellar velocity is the mean velocity of the two peaks, and the expansion velocity of the envelope is half of their velocity difference. In some cases, the two OH peaks are highly asymmetric. Single-peaked OH spectra could correspond to cases of asymmetric intensity, where one of the peaks falls below the sensitivity of the observations. { Thus, it is highly unlikely that the single OH maser feature reported for this source (at $V_{\rm LSRK}=-118$ km s$^{-1}$) may trace the stellar velocity.}

{ We note that \cite{dea04} listed a second, weaker OH component at $-104.8$ km s$^{-1}$. This component can be seen in their published spectrum, and it is also hinted in the spectrum shown by \cite{sev97}, although the latter do not mention this possible component. If this were real, then the stellar velocity could be $\simeq -111$ km s$^{-1}$, with an expansion velocity in its circumstellar envelope of $\simeq 6.6$ km s$^{-1}$. We have downloaded from the ATCA archive the interferometric OH data taken by \cite{sev97}, and imaged them with different weighting of the visibilities (robust parameter = $-2$, 0.5, and 2 in {\sc miriad}). While a component peaking at $V_{\rm LSRK} \simeq -117.1$ km s$^{-1}$ is present in the map regardless of the weighting, a  weaker emission in a single channel (at $-102.6$ km s$^{-1}$) is tentatively detected only in the map with robust parameter 0.5, and at a position $\simeq 5$ arcsec away from the first component and the stellar position. Since masers are unresolved by ATCA, their detection should not depend on the particular weighing used. This, and the apparent angular separation indicates that the OH emission at $-102.6$ km s$^{-1}$, if real, is probably diffuse interstellar emission, not associated with IRAS 15544-5332. Thus, only one OH maser component is confirmed toward the source, and it cannot be used to estimate the stellar velocity.}

{ \cite*{van00} estimated a heliocentric stellar velocity of $-68$ km s$^{-1}$ (which corresponds to $V_{\rm LSRK} \simeq -67$ km s$^{-1}$) using Br$\gamma$ spectra, and quoted an uncertainty of $\simeq 10$ km s$^{-1}$. The velocity of the water maser emission is blueshifted by only 6 km s$^{-1}$ with respect to that velocity, which would argue against a WF nature for IRAS 15544-5332. However, the Br$\gamma$ obtained by \cite{van00} has a low spectral resolution ($\lambda/\Delta\lambda = 2000$, corresponding to a velocity resolution of $\simeq 150$ km s$^{-1}$) and it is possible that there are different unresolved features that preclude a precise determination of the velocity. Moreover, 
there is an absorption feature immediately adjacent to the Br$\gamma$ emission line that was fitted to obtain the velocity estimate. This absorption feature could remove part of the emission, resulting in an apparent velocity significantly shifted with respect to the actual stellar velocity.

The relative velocities of the OH and water maser emission can give us some useful hints about the stellar and expansion velocities of the envelope. As mentioned above, OH emission in evolved stars typically traces the approaching and receding sides of the circumstellar envelope. However, in the case of water masers \citep{eng96,dea07}, two different cases are possible, depending on the type of source. The emission is close to the stellar velocity in Mira-type stars, while in more massive OH/IR stars the spectra are double-peaked (as in the case of OH masers), with velocities close, but slightly lower with respect to the stellar velocity than those of OH. 

If we assume that the water maser emission in IRAS 15544-5332 is close to the stellar velocity (Mira-type star), then the OH peak traces an expansion velocity of $\simeq 45$ km s$^{-1}$. This would be too large for AGB stars, especially for Mira-type objects, which present low expansion velocities \citep[5-10 km s$^{-1}$,][]{dea07}}

{ On the other hand, if IRAS 15544-5332 is a more massive OH/IR star, } the observed OH peak 
could trace to the blueshifted half of the envelope, while the H$_2$O peak would trace the 
redshifted one. This would correspond to a velocity difference between the two halves of the envelope of 
$\simeq 40$ km s$^{-1}$ { (expansion velocity $\simeq 20$ km s$^{-1}$)}, 
similar to the difference seen between the OH peaks in evolved stars (see, e.g., Fig. \ref{fig:spec_i15oh1612}). { The stellar velocity in this case would be $\simeq -96$ km s$^{-1}$. This is the scenario that we favour for this source}.

{ In either case (water maser at the stellar velocity or offset from it)   there is no confirmed high-velocity H$_2$O emission in this source, and it could certainly be a normal AGB or post-AGB star}. We conclude that there is no evidence, either in our data or in the literature, that IRAS 15544-5332 may be a WF. Thus, it should  not be considered as a WF candidate.

\subsection{IRAS 17291-2147}

\label{sec:17}

\begin{figure}
	\includegraphics[width=0.9\columnwidth]{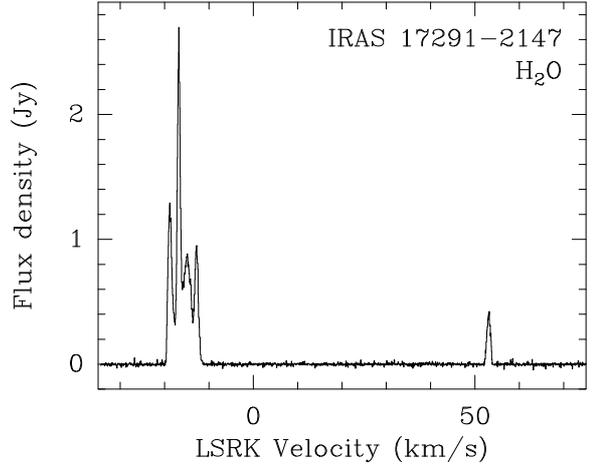}
	\caption{Water maser spectrum toward IRAS 17291-2147.}
	\label{fig:spec_i17h2o}
\end{figure}

This source was suggested as a possible WF by \cite{gom15b}, based on a single-dish water-maser spectrum that showed a total velocity spread of $\simeq 96$ km s$^{-1}$.  Our interferometer observations show at least four blueshifted and one redshifted component, spanning a velocity range of $\simeq 74$ km s$^{-1}$ (Fig. \ref{fig:spec_i17h2o} and Table \ref{tab:h2o}). No OH emission was detected (Table \ref{tab:oh}). { Non-detection of OH emission at 1612 MHz was also reported by \cite{tel91}.}

\begin{figure}
	\includegraphics[width=0.9\columnwidth]{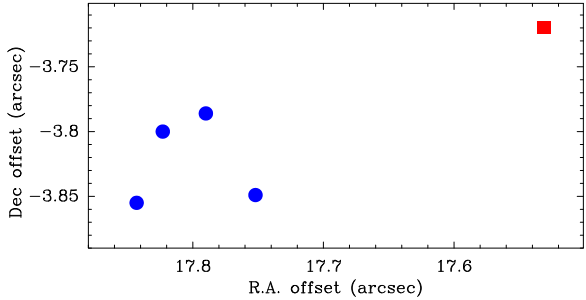}
	\caption{Location of water maser components in IRAS 17291-2147. Blue circles represent the components with $V_{\rm LSRK} < -12$ km s$^{-1}$, and the red square is the component at +53.21 km s$^{-1}$. The axis coordinates are arcsec offsets with respect to the phase centre of the observations (shown in Table \ref{tab:obs}, and which corresponds to the position given in the IRAS point source catalog). }
	\label{fig:map_i17h2o}
\end{figure}

All water maser components seem to arise from IRAS 17291-2147 (Table \ref{tab:i17_h2o}). Their location is $\simeq 3$ arcsec away { from the IR counterpart  of the source (2MASS J17321117-2150022, WISEA J173211.18-215002.3, AKARI-IRC-V1 J1732111-215002)}.  There is no other IR source within { 9 or 15 }arcsec of the maser emission in the 2MASS or WISE catalogs, { respectively, }so this { isolated} IR source is the most likely pumping source of the masers. This also indicates that the absolute positional uncertainty of our water maser observations could be $\simeq 3$ arcsec. The blueshifted maser components ($V_{\rm LSRK} < -12$ km s$^{-1}$) seem to cluster together, with the redshifted component ($V_{\rm LSRK} = +53.21$) being located $\simeq 0.3"$ to the west-northwest (Fig. \ref{fig:map_i17h2o}). {  Since there is only one redshifted component, any conclusion should be taken with care. However, the fact that this redshifted component is significantly separated both spatially and spectrally from the other maser components, 
suggests that the maser emission may be tracing} a bipolar jet, with an orientation of $\simeq -70^\circ$, and confirms the object as a WF. 
{ If we assume that the central star is located between the red and blueshifted maser emission, we would interpret these masers as tracing two lobes of a bipolar jet, and the stellar velocity would be $\simeq 18$ km s$^{-1}$. However, we cannot discard the possibility that one of the maser groups is located close to the central star both in velocity and position, and therefore, the maser emission is tracing only one lobe of a jet.}
Observations with higher angular resolution would provide a better characterization of { this possible} maser jet.

\begin{table}
	\centering
	\caption{Water maser components in IRAS 17291-2147.}
	\label{tab:i17_h2o}
	\begin{tabular}{lllll} %
		\hline
$V_{\rm LSRK}$ & $S_\nu$  & R.A.(J2000) & Dec(J2000) & Error \\
(km s$^{-1}$)  & (Jy) &						& & (mas) \\
		\hline
$-18.81$ & $1.276\pm 0.008$ & 17:32:11.381 & $-21$:50:02.85 & 30\\ 
$-16.77$ & $2.656\pm 0.011$ & 17:32:11.3749 & $-21$:50:02.846 & 23\\ 
$-14.86$ & $0.871\pm 0.008$ & 17:32:11.378 & $-21$:50:02.78 & 50 \\
$-12.75$ & $0.937\pm 0.008$ & 17:32:11.380 & $-21$:50:02.80 & 40\\
$+53.21$ & $0.413\pm 0.009$ & 17:32:11.359 & $-21$:50:02.72 & 90\\
		\hline
	\end{tabular}
\end{table}

\subsection{IRAS 18596+0315 (OH 37.1-0.8)}


Previous observations of water maser emission toward this source with single-dish telescopes \citep*{eng86,bra94,eng02,dea07} showed a highly variable spectrum, with multiple spectral components spanning a velocity range of $\simeq 57$ km s$^{-1}$, significantly larger than the velocity covered by the double-peaked OH spectrum \citep[$\simeq 28$ km s$^{-1}$; see, e.g.,][]{sev01}. Interferometric water maser observations \citep{gom94} revealed only a spectral component, which was found to be coincident with the OH emission within $1''$. {  The velocity of this water maser component is outside the range of the OH maser emission, which strongly suggests that this object is a water fountain}. There is however, no reported interferometric data showing the presence of water maser emission spanning a large velocity range, {  as suggested by the single-dish data}.

\begin{figure}
	\includegraphics[width=0.9\columnwidth]{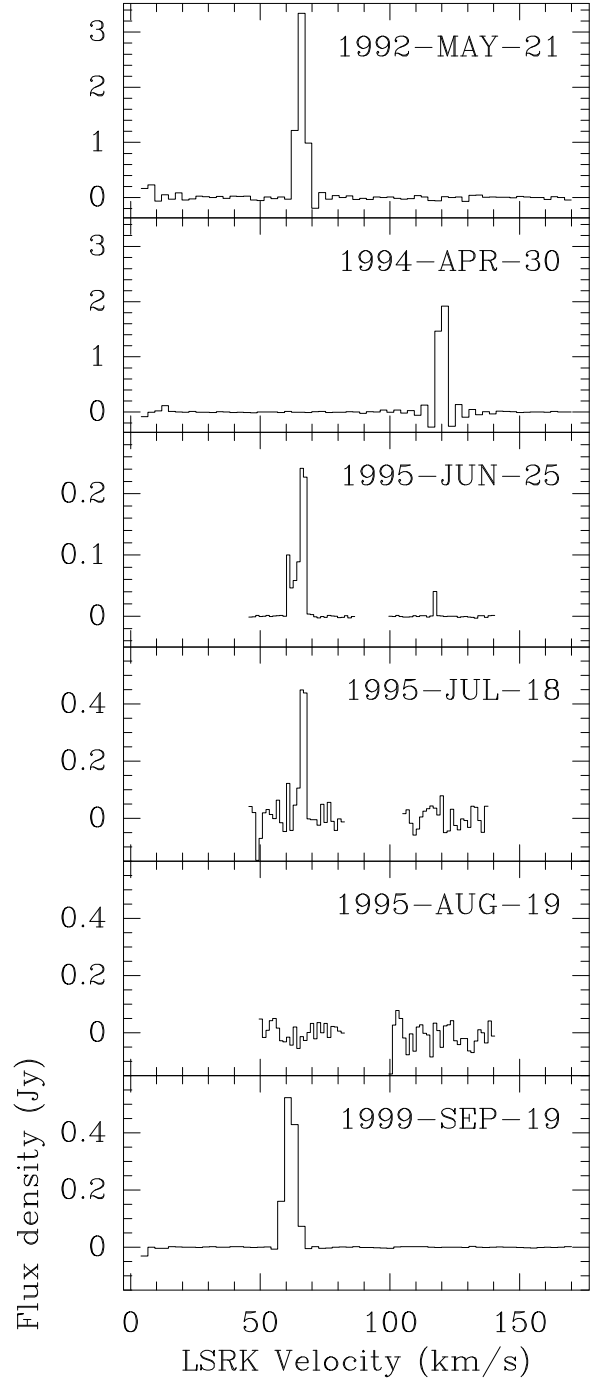}
	\caption{Water maser spectra toward IRAS 18596+0315, obtained from VLA archival data. }
	\label{fig:spec_i18vla}
\end{figure}

\begin{figure}
	\includegraphics[width=0.9\columnwidth]{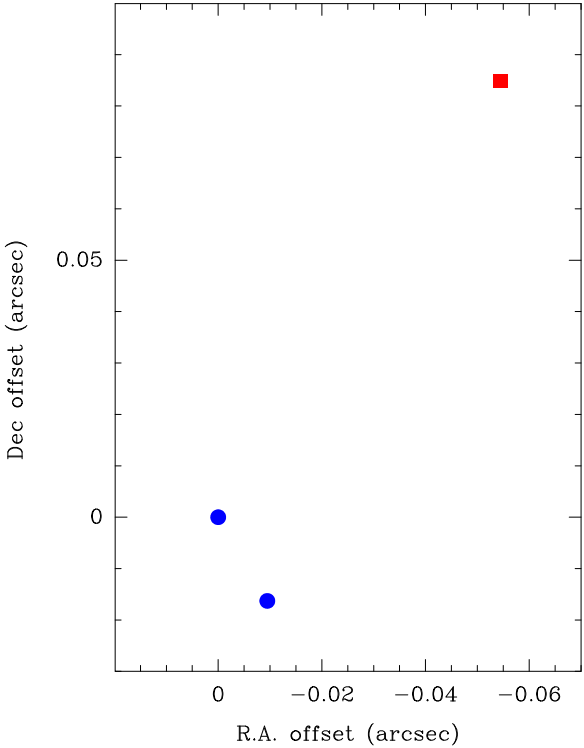}
	\caption{Location of water masers in IRAS 18596+0315 detected on 1995-Jun-25. Blue circles represent the components with $V_{\rm LSRK} < +67$ km s$^{-1}$, and the red square is the component at +117.4 km s$^{-1}$. The axis coordinates are arcsec offsets with respect to the strongest maser component detected on that date (see Table \ref{tab:i18vla}).}
	\label{fig:map_i18h2o}
\end{figure}

We do not detect any water maser emission in our ATCA data (Table \ref{tab:h2o}). However, we have processed archival data from all water maser observations of this source carried out with the VLA. The resulting spectra are shown in Fig. \ref{fig:spec_i18vla}. The data taken in 1992 May 21 are the ones published by \cite{gom94}, while the rest are unpublished. The spectrum shows high variability, and only on 1995 June 25 it showed more than one component. In all cases, the spectral components arise from IRAS 18596+0315. In particular, that the three components observed on 1995 June 25 are associated with this source confirm its WF nature. The components detected on 1995-Jun-25 seem to be distributed in a bipolar pattern (Fig. \ref{fig:map_i18h2o}), extending for $\simeq 0.1''$ along a position angle $\simeq -30^\circ$. { However, with only three components, the characteristics of a putative jet are highly uncertain}.  The position of the centre of this water maser distribution is $\la 0.25$ arcsec away from the { position of  an IR} source { detected by Spitzer and WISE (SSTSL2 J190206.28+032015.6, WISEA J190206.28+032015.7), which seems to be its most likely IR counterpart.  No conterpart in the 2MASS catalog was detected. The source is relatively isolated, since other sources in the Spitzer, WISE and Akari catalogs are $> 12$ arcsec away.}

\begin{table*}
	\centering
	\caption{Parameters of archival water maser data on IRAS 18596+0315.}
	\label{tab:i18vla}
	\begin{tabular}{llllllll} %
		\hline
Date & $V_{\rm LSRK}$  & R.A.(J2000) & Dec(J2000) & Error & $S_\nu$ & $\theta_{\rm fwhm}$ & $\theta_{\rm pa}$\\
	& (km s$^{-1}$)	  & 			&				& (mas) & (Jy) & (arcsec) & (deg)\\
		\hline
1992-May-21 & $+66\pm 3$ & 19:02:06.289 & $+03$:20:15.71 & 30 & $3.303\pm 0.015$ & $1.2\times 1.0$ & $-31$\\
1994-Apr-30 & $+121\pm 3$ & 19:02:06.283320& $+03$:20:15.82196 & 0.14 & $1.852\pm 0.007$ & $0.08\times 0.07$ & $-7$ \\ 
1995-Jun-25 & $+60.7\pm 1.3$ &  19:02:06.28618 & $+03$:20:15.8116 & 0.5 & $0.0981\pm 0.0011$ & $0.10\times 0.09$ & $-13$\\ 
			& $+66.0\pm 1.3$ & 19:02:06.286812 & $+03$:20:15.82790 & 0.20 & $0.2381\pm 0.0010$ & \\
			& $+117.4\pm 1.3$ & 19:02:06.28315 & $+03$:20:15.9125 & 1.2 & $0.0389\pm 0.0010$\\
1995-Jul-18 & $+67.3\pm 1.3$ & 19:02:06.2863 & $+03$:20:15.822 & 3 & $0.49\pm 0.03$ & $0.10\times 0.08$ & $-30$ \\ 
1995-Aug-19 & 				&				&					& &$<0.03$ & $0.10\times 0.09$ & $-24$\\
1999-Sep-19 & $+61\pm 3$ & 19:02:06.274264 & $+03$:20:15.94078 & 0.09 & $0.8803\pm 0.0013$ & $0.18\times 0.09$ & $+44$\\
		\hline
	\end{tabular}
\end{table*}

\begin{figure}
	\includegraphics[width=0.9\columnwidth]{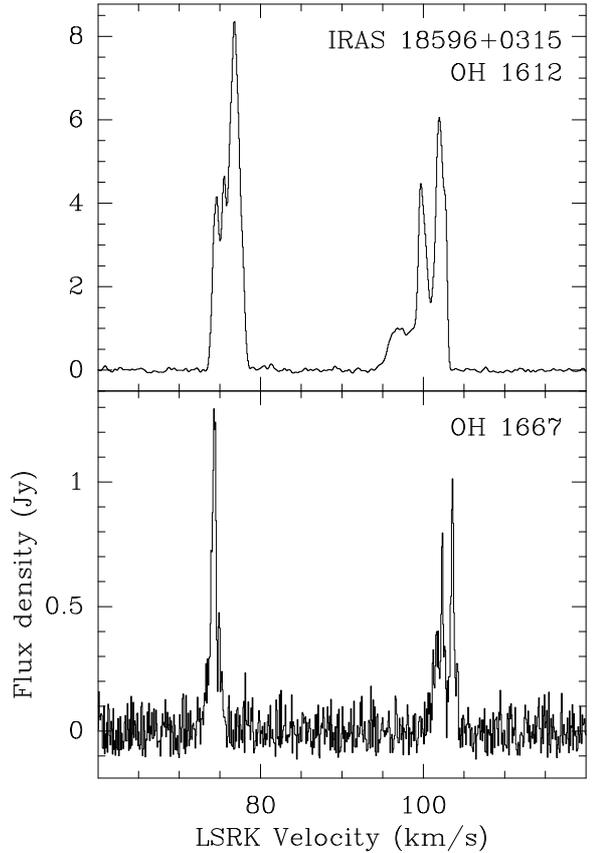}
	\caption{OH maser spectum at 1612 and 1667 MHz toward IRAS 18596+0315.}
	\label{fig:spec_i18oh}
\end{figure}

Our OH observations of this source have poor uv coverage, and we could not { obtain precise information about the location of the emission}. We obtained spectra using task uvspec of {\sc Miriad}. They show that 
 OH emission was detected both at 1612 and 1667 MHz (Fig. \ref{fig:spec_i18oh}), showing multiple components distributed in two groups, with a total velocity spread of $\simeq 30$ km s$^{-1}$, These OH spectra are similar to the ones published by \cite{wol12}, although the blueshifted emission at 1667 MHz is a factor of $\simeq 3$ stronger in our data. Figs. \ref{fig:spec_i18vla} and \ref{fig:spec_i18oh} clearly show that the H$_2$O emission is well outside the velocity range covered by OH masers, further confirming that this source is a WF.

Interferometric observations of OH maser emission at 1612 MHz toward this source \citep*{ami11} show that blue and redshifted components are spatially separated by $\simeq 100$ mas { roughly} along the E-W direction. 
{ A comparison with the water maser distribution is not straightforward, given the much higher positional accuracy of the OH data, and the limited number of spectral water maser components. In the models of \cite{ami11}, we would expect that the red/blue OH lobes are oriented along the same direction as that of a water maser jet, regardless of whether the OH traces a jet or an equatorial (expanding) torus. However, if we take our determination of the orientation of the water maser structure at face value, that would not be the case in this source. A more precise comparison between water and OH masers would require VLBI observations of the water maser emission, where the jet direction could be better determined with higher positional accuracy and/or with a measurement of proper motions.}

\subsection{IRAS 19067+0811 (V1368 Aql, OH 42.3-0.1)}


Single-dish spectra of this source between 1976 and 1987 showed water maser emission with a redshidfted component at 78 km s$^{-1}$ slightly outside (by $\simeq 3$ km s$^{-1}$) the velocity range of OH emission \citep{eng86,nym86,eng96}. Using a wider bandwidth, \citep{gom94} also detected an additional high-velocity blueshifted component at 10 km s$^{-1}$, giving a separation of  $\simeq 70$ km s$^{-1}$  between both components, suggesting that this source was a WF candidate. However, later single-dish  \citep{kim10,vle14} and interferometric observations \citep{gom94,vle14} failed to detect any high-velocity emission and, in particular, the blueshifted component at 10 km s$^{-1}$ was absent. 

We did not detect any water maser emission toward this source. Therefore, its nature as a WF is not yet confirmed by any interferometric observation. So far, there is no evidence that the high-velocity water maser emission detected by \cite{gom94} actually arises from IRAS 19067+0811.

\section{Discussion}

\subsection{The nature of the confirmed objects}

{ IRAS 17291-2147 and IRAS 18596+0315 show water maser components spanning a large velocity range, and all arising from these sources. This would confirm them as WFs, provided that they are indeed evolved objects. Objects in early evolutionary stages, such as high-mass young stellar objects (YSOs) and H {\sc ii} regions can have infrared colours similar to those of AGB and post-AGB stars \citep{lum02,sua06}, and these young objects can also have water maser components with large velocities \citep[see, e.g.,][]{wal82}. 

Our non-detection of radio continuum emission rules out that these objects are H {\sc ii} regions, since our continuum sensitivity at 23 GHz (Section \ref{sec:results}) would be enough to detect any galactic H {\sc ii} region \citep{urq09}. Massive YSOs in early stages may not have yet developed an H {\sc ii} region, but radio continuum from ionized jets or shocks is common in these objects \citep[e.g.][]{pur16}. Our continuum results suggest that IRAS 17291-2147 and IRAS 18596+0315 are indeed evolved objects. This is further confirmed by the archival Spitzer and WISE images, which show point-like objects  dominating the emission at the longest wavelengths, and without associated filaments and nebulosities usually present in massive star-forming regions \citep{urq09}. Moreover, their SEDs peak at $\simeq 25$ $\mu$m \citep{rl09,yun17}, which is typical of evolved stars, rather than of YSOs \citep{urq09}.
}

\subsection{The group of confirmed water fountains}

In this paper we have confirmed two sources (IRAS 17291-2147 and IRAS 18596+0315) as WFs. The other two remain unconfirmed. IRAS 19067+0811 still requires a confirmation that the high-velocity components detected with single-dish telescopes actually arise from the source, and not from other nearby sources. In the case of IRAS 15544-5332, there is no real evidence in the literature that it is a WF, and its identification as a candidate was based on questionable arguments (Sec. \ref{sec:15}).

The number of known WFs is small, but there is still some confusion in the literature with regard to which sources can be considered as confirmed examples of this class. We emphasize that the mere presence of apparent high-velocity water maser features in a single-dish spectrum is not enough to classify a source as a bona fide WF, since the large beam could pick up emission from different sources (yielding a spectrum with a falsely large velocity spread) or from a single but unrelated object, such as high-mass young stellar object in the neighbourhood \citep[as in the case of IRAS 19071+0857,][]{gom15b}. Moreover, interferometric observations can provide accurate positional information to identify whether the maser emission is tracing a jet.

To summarize the situation to date, Table \ref{tab:confirmed} lists the WFs that have been confirmed with interferometric observations, including the two in this paper. This table shows the total velocity span observed in the water masers in these sources, and the literature references for their confirmation as WFs. { Our discussion below regarding the global characteristics of WFs is based on this group of bona fide WFs.}

\begin{table}
	\centering
	\caption{Confirmed water fountains { to date}.}
	\label{tab:confirmed}
	\begin{tabular}{llrl} %
		\hline
IRAS name & Alternate name  & $V_{\rm range}$ & References  \\
&&(km s$^{-1}$)  &					\\
		\hline
    15103-5754 & GLMP 405 & 70  & 1\\
    15445-5449 & OH 326.5-0.4 & 90 &  2\\
   	16342-3814 & OH 344.1+5.8 & 260 &  3\\
   	16552-3050 & GLMP 498  & 170 & 4  \\
   	17291-2147 &			& 70 & 5\\
   	18043-2116 & OH 9.1-0.4 & 400 & 6 \\
   	18113-2503 & PM 1-221 & 500 &  7\\
   	18139-1816 & OH 12.8-0.9 & 50 & 8\\
   	18286-0959 & OH 21.80-0.13 & 200 &  9\\
   	18450-0148 & W43A & 180 &  10\\
   	18455+0448 & & 40 &  11\\
   	18460-0151 & OH 31.0-0.2 & 310 &  12\\
   	18596+0315 & OH 37.1-0.8 & 60 & 5\\
   	19134+2131 & & 100 &  13\\
   	19190+1102 & PM 1-298 & 100 & 14 \\
		\hline
	\end{tabular}\\
	References: 1: \cite{gom15a}. 2: \cite{per11}. 3: \cite{cla09}. 4. \cite{sua08}. 5: This paper. 6: \cite{wal09}. 7: \cite{gom11}. 8: \cite{bob05}. 9: \cite{yun11}. 10: \cite{ima02}. 11: \cite{vle14}. 12: \cite{ima13}. 13: \cite{ima04}. 14: \cite{day10}
\end{table}

\subsection{OH emission in water fountain stars}

Both IRAS 17291-2147 and IRAS 18596+0315 show 
{  maser components that are well separated, both spectrally and spatially, and that are consistent with tracing collimated jets}. 
We also note that IRAS 18596+0315 harbours OH maser emission, whereas IRAS 17291-2147 does not. All WFs have been observed in OH, and only in the case of IRAS 16552-3050, IRAS 18113-2503, and IRAS 19134+2131 (in addition to IRAS 17291-2147) this emission has not been detected \citep*{lew87,hu94,yun14}. The remaining WFs tend to show double-peaked OH spectra similar to other AGB and post-AGB stars, probably tracing the expansion of the remnant circumstellar envelope. However, the spatial distribution of OH in some sources is not compatible with a spherically symmetric envelope \citep[e.g.,][]{sah99,ami11}. It is interesting to consider the possible reasons of this difference among WFs, regarding the presence or not of OH maser emission. A possibility is that the circumstellar envelope in some WFs is disrupted in such a way that the velocity coherence along the line of sight is lost, thus suppressing OH maser emission. A companion star could induce such a disruption \citep{lew87}. { However}, the presence of collimated jets suggests the presence of binary systems \citep{sok14} in all WFs, { but this does not prevent the excitation of OH masers in some of them}. More complex interactions, such as multiple stellar systems \citep{sok16} could give rise to different velocity fields, breaking the velocity coherence in the envelope. 


\subsection{Velocity spread of water masers}

As mentioned in the introduction, WFs are identified by a velocity range of their water maser emission 
which is larger than that of the OH emission. However, Table \ref{tab:confirmed} reveals a very large
spread in the velocity range of the water masers in the confirmed WFs, from $\simeq 40$ up to 500 km s$^ {-1}$. These imply jet velocities projected along the line of sight of $\simeq 20$ to 250 km s$^ {-1}$.
These velocity ranges of water masers may depend on several factors. An obvious one is the different orientation of the jet axis with respect to the observer. Another one is the source age, if there is an evolution of the energetics of mass-loss of WFs with time. 

To investigate possible trends, we have retrieved the photometric data from 2MASS \citep{skr06} and
WISE \citep{wri10} archives of the WFs in Table \ref{tab:confirmed}, and calculated different colour indices as a measure of extinction. We could expect that in sources whose jets move near the plane of the sky, the central stars would be more obscured than those with jets along the line of sight, if jets clear up cavities in the circumstellar envelope. On the other hand, more evolved sources are also expected to show lower extinction, since the density of their envelopes will decrease as it expands, and jets would have had more time to clear up cavities. However, no correlation was found between infrared colour indices and the observed velocity ranges.

Another possible indicator of age is the spatial distribution of water masers. { The maser emission in some WFs exclusively trace collimated jets, whereas in others \citep[e.g., W43A or IRAS 18460-0151][]{ima02,ima13} some spectral components also trace equatorial ejections. The reason why not all WFs show equatorial structures traced by their water maser emission is not clear. Considering that in more evolved objects (PNe), water maser emission mostly traces equatorial structures rather than jets \citep{gom08}, we speculate that WFs showing only masers from a jet are relatively younger than those presenting equatorial ejections also.}
 However, we have not seen any correlation of the spatial distribution with maser velocities nor with infrared colours.

\subsection{Infrared characteristics of water fountain stars and other obscured post-AGB objects}

\begin{figure}
	\includegraphics[width=\columnwidth]{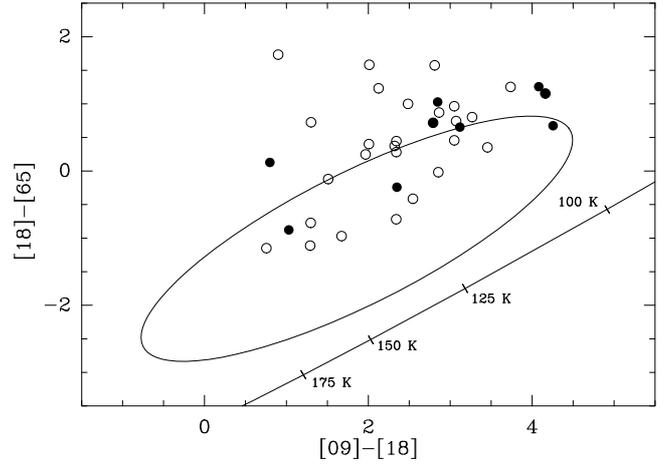}
	\caption{ Akari colour-colour diagram of water fountains and obscured post-AGB candidates. The axes are differences in magnitudes between Akari bands (9, 18, and 65 $\mu$m). Filled circles represent confirmed water fountains. Open circles represent obscured post-AGB candidates in Ramos-Larios et al. (2009, 2012).
	 The solid line represents the locii of the colours for black-body brightness distributions. The ellipse
	  is the area marked by Yung et al. (2014) as encompassing most of the post-AGB stars
	   in the Meixner et al. (1999) sample.}
	\label{fig:colorakari}
\end{figure}

{ The infrared characteristics of WFs may shed some light on whether these objects differ from other post-AGB stars. \cite{yun14} suggested that the location of post-AGB stars on a colour-colour diagram of Akari magnitudes may be correlated with temperature, as they tend to be distributed in that diagram on an area elongated paralell to the expected colour of a black body (the ellipse in Fig. \ref{fig:colorakari}). Their WF candidates seemed to be at the lowest temperature end of the post-AGB region. We have extended this diagram in Fig. \ref{fig:colorakari} by including all confirmed water fountains listed in Table \ref{tab:confirmed}. As a comparison group, we have plotted in that figure the obscured post-AGB stars from \citet[][2012]{rl09}, instead of the post-AGB stars used in \cite{yun14}, which were a sample compiled by \cite{mei99}, most of which are optically visible objects. Since WFs tend to be optically obscured, a comparison with more similar objects will help to isolate other differential characteristics. In this plot, we see that water fountains are spread over a larger area than assumed { by} \cite{yun14}, and that they do not show an evident different distribution from that of obscured post-AGB stars. We also note that the location of obscured post-AGBs in that diagram is displaced from that of optically visible ones, and that the former do not show a clear correlation with temperature in that diagram.
}

\begin{figure}
	\includegraphics[width=\columnwidth]{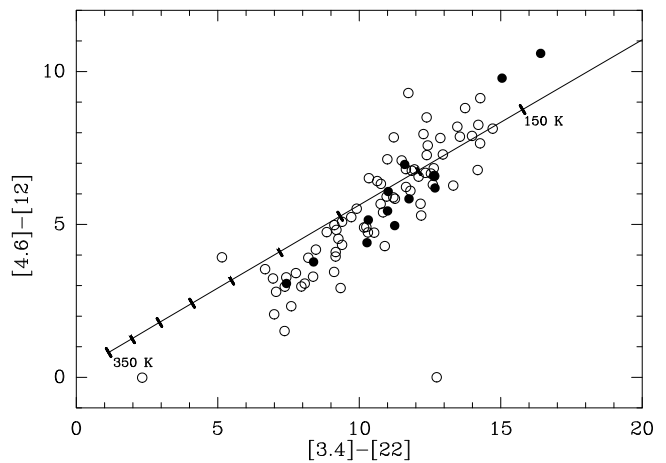}
	\caption{{ WISE} colour-colour diagram of water fountains and obscured post-AGB candidates. The axes are differences in magnitudes between WISE bands {  (3.4, 4.6, 12, and 22 $\mu$m)}. Filled circles represent confirmed water fountains. Open circles represent obscured post-AGB candidates in Ramos-Larios et al. (2009, 2012). { The solid line represents the locii of the colours for black-body brightness distributions. The tick marks in the black-body line go from 150 to 350 K at increment steps of 25 K.}}
	\label{fig:colorwise}
\end{figure}

{ In Fig. \ref{fig:colorwise} we present a colour-colour diagram with magnitudes from the WISE archive.} 
{ This diagram is more clearly correlated with temperature, since obscured post-AGBs and WFs are located close to the expected colours of a black body. } A Kolmogorov-Smirnov test did not show any difference in their distribution in that diagram among WFs and obscured post-AGB stars. 
{ Figs. \ref{fig:colorakari} and \ref{fig:colorwise} suggest} that WFs do not have any distinctive characteristic from other obscured post-AGBs, apart from the detection of high-velocity maser emission. That some obscured post-AGBs do not show this high-velocity emission might be due to variability, if the WF phenomenon is episodic. Alternatively, effects such as the orientation of the jet with the line of sight could be relevant, as maser emission is highly beamed. 
{ In summary, these results support the idea suggested by \cite{yun17} that WF jets do not necessarily represent the earliest collimated mass loss processes in evolved stars.}

{ Apart from the absence of any difference between the distributions of WFs and post-AGB stars in Fig. \ref{fig:colorwise}, it is interesting to see the trends in this figure. While the general trend is linear, and close to the line traced by the different temperatures in a black-body energy distribution, there are some systematic deviations from the { latter}, most notably that the slope of the source distribution is steeper than the black body line. This is not surprising, since the spectral energy distribution (SED) of these objects cannot be described by a single blackbody \citep{yun17}. So while the position of a source in this diagram may be correlated with temperature, it would not give a precise value of this temperature. The ``coldest'' WFs in Fig. \ref{fig:colorwise} (the ones with a larger value of [3.4]-[22]) are IRAS 16552-3050 and W43A, while the ``warmest'' ones are OH 12.8-0.9 and IRAS 18455+0448. The SED of the two ``coldest'' sources \citep[Fig. 3 in][]{yun17} rises sharply between 3.4 and 22 $\mu$m, since it peaks at $\lambda \ga 22$  $\mu$m, as in the case of black bodies with $T\le 150$ K. On the other hand, the SED of the two ``warmest'' objects \citep{yun17} does not change much between 3.4 and 22 $\mu$m. Therefore, their colours in Fig. \ref{fig:colorwise} mimic a blackbody also peaking within that range ($T\simeq 225$ K). The colours in Fig. \ref{fig:colorwise} may be related with age and progenitor mass, but these dependencies are not easy to disentangle, in principle. Note, for instance, that AGB stars tend to have SED peaks at shorter wavelengths than post-AGB stars \citep{yun17}, so they would appear to the left and bottom in this figure. Based on their SEDs and the relatively low velocities of their water masers (compared with other WFs), \cite{yun17} suggested that OH 12.8-0.9 and IRAS 18455+0448 could be among the youngest WFs. However, note that W43A, the only WF that has been clasified as an AGB star, is at the opposite end of the plot. An alternative scenario is that objects toward the ``warm'' areas of this diagram have a contribution in their SEDs from a hotter structure than the envelope, such as a circumstellar disk. An accurate modeling is necessary to properly interpret the origin of the colour variations in Fig. \ref{fig:colorwise}, which may make this diagram a useful diagnostic tool to classify WFs.}



\section{Conclusions}

We presented interferometric data from new ATCA observations and from the VLA archive toward four WF candidates. Our conclusions are as follows:

\begin{itemize}
\item We confirm IRAS 17291-2147 and IRAS 18596+0315 (OH 37.1-0.8) as bona fide members of the WF class, with high-velocity water maser emission {  consistent with} tracing bipolar jets. With these, there are so far 15 confirmed WFs. 
\item We see no evidence in our data nor in the literature that IRAS 15544-5332 is a WF. This object should not be considered a WF candidate.
\item We failed to detect any water maser emission towards IRAS 19067+0811. Its nature as a WF remains to be confirmed.
\item We do not see any significant trend relating the velocity spread in WFs with indicators of different ages or jet orientations.
\item The infrared colours of WFs are not significantly different from other obscured post-AGB stars. That some of the later do not show high-velocity water maser emission could be a variability or orientation effect, rather than reflecting a fundamental physical difference with WFs. 
\end{itemize}

\section*{Acknowledgements}

The Australia Telescope Compact Array  is part of the Australia Telescope National Facility, which is funded by the Australian Government for operation as a National Facility managed by CSIRO. The National Radio Astronomy Observatory is a facility of the National Science Foundation operated under cooperative agreement by Associated Universities, Inc.
 J.F.G. and L.F.M are
partially supported by Ministerio de Econom\'{\i}a, Industria y Competitividad (Spain) grant  AYA2014-57369-C3-3 (co-funded by
FEDER). JFG is also supported by Ministerio de Educaci\'on, Cultura y Deporte (Spain), under the mobility program for senior scientists at foreign universities and research centres. L.U. acknowledges support from PRODEP (Mexico).
This publication makes use of data products from the Two Micron All Sky Survey (which is a joint project of the University of Massachusetts and the Infrared Processing and Analysis Center/California Institute of Technology, funded by the National Aeronautics and Space Administration and the National Science Foundation), data products from the Wide-field Infrared Survey Explorer (which is a joint project of the University of California, Los Angeles, and the Jet Propulsion Laboratory/California Institute of Technology, funded by the National Aeronautics and Space Administration), and observations with AKARI (a JAXA project with the participation of ESA). 





\begin{thebibliography}{99}

\bibitem[\protect\citeauthoryear{Amiri, Vlemmings, \& van Langevelde}{Amiri et al.}{2011}]{ami11} Amiri N., Vlemmings W., van Langevelde H.~J., 2011, A\&A, 532, A149 


\bibitem[\protect\citeauthoryear{Boboltz \& Marvel}{2005}]{bob05} Boboltz D.~A., Marvel K.~B., 2005, ApJ, 627, L45 

\bibitem[\protect\citeauthoryear{Brand et al.}{1994}]{bra94} Brand J., et al., 1994, A\&AS, 103,  541

\bibitem[\protect\citeauthoryear{Caswell}{1998}]{cas98} Caswell J.~L., 1998, MNRAS, 297, 215 

\bibitem[\protect\citeauthoryear{Caswell \& Haynes}{1975}]{cas75} Caswell J.~L., Haynes R.~F., 1975, MNRAS, 173, 649 


\bibitem[\protect\citeauthoryear{Claussen, Sahai, \& Morris}{2009}]{cla09} Claussen M.~J., Sahai R., Morris M.~R., 2009, ApJ, 691, 219 

\bibitem[\protect\citeauthoryear{Condon}{1997}]{con97} Condon J.~J., 1997, PASP, 109, 166 

\bibitem[\protect\citeauthoryear{Condon et al.}{1998}]{con98} Condon J.~J., Cotton W.~D., Greisen E.~W., Yin Q.~F., Perley R.~A., Taylor G.~B., Broderick J.~J., 1998, AJ, 115, 1693 

\bibitem[\protect\citeauthoryear{Day et al.}{2010}]{day10} Day F.~M., Pihlstr{\"o}m Y.~M., Claussen M.~J., Sahai R., 2010, ApJ, 713, 986 


\bibitem[\protect\citeauthoryear{Deacon, Chapman, \& Green}{Deacon et al.}{2004}]{dea04} Deacon R.~M., Chapman J.~M., Green A.~J., 2004, ApJS, 155, 595 

\bibitem[\protect\citeauthoryear{Deacon et al.}{2007}]{dea07} Deacon R.~M., Chapman J.~M., Green A.~J., Sevenster M.~N., 2007, ApJ, 658, 1096 

\bibitem[\protect\citeauthoryear{Desmurs}{2012}]{des12} Desmurs, J.-F.\ 2012, IAU Symp. 287, Cosmic Masers from OH to H0, ed.
R.~S. Booth, E.~M.~L. Humphreys, \& W.~H.~T. Vlemmings (Cambridge:
Cambridge Univ. Press), 217


\bibitem[\protect\citeauthoryear{Engels}{2002}]{eng02} Engels D., 2002, A\&A, 388, 252 

\bibitem[\protect\citeauthoryear{Engels \& Lewis}{1996}]{eng96} Engels D., Lewis B.~M., 1996, A\&AS, 116, 117 


\bibitem[\protect\citeauthoryear{Engels, Schmid-Burgk, \& Walmsley}{Engels et al.}{1986}]{eng86} Engels D., Schmid-Burgk J., Walmsley C.~M., 1986, A\&A, 167, 129 

\bibitem[\protect\citeauthoryear{Gaylard \& Whitelock}{1988}]{gay88} Gaylard M.~J., Whitelock P.~A., 1988, MNRAS, 235, 123 

\bibitem[\protect\citeauthoryear{G{\'o}mez et al.}{2008}]{gom08} G{\'o}mez J.~F., Su{\'a}rez O., G{\'o}mez Y., Miranda L.~F., Torrelles J.~M., Anglada G., Morata {\'O}., 2008, AJ, 135, 2074 

\bibitem[\protect\citeauthoryear{G{\'o}mez et al.}{2011}]{gom11} G{\'o}mez J.~F., Rizzo J.~R., Su{\'a}rez O., Miranda L.~F., Guerrero M.~A., Ramos-Larios G., 2011, ApJ, 739, L14 

\bibitem[\protect\citeauthoryear{G{\'o}mez et al.}{2015a}]{gom15a} G{\'o}mez J.~F., et al., 2015a, ApJ, 799, 186 

\bibitem[\protect\citeauthoryear{G{\'o}mez et al.}{2015b}]{gom15b} G{\'o}mez J.~F., Rizzo J.~R., Su{\'a}rez O., Palau A., Miranda L.~F., Guerrero M.~A., Ramos-Larios G., Torrelles J.~M., 2015b, A\&A, 578, A119 

\bibitem[\protect\citeauthoryear{G{\'o}mez et al.}{1994}]{gom94} G{\'o}mez Y., Rodr{\'{\i}}guez L.~F., Contreras M.~E., Moran J.~M., 1994, RMxAA, 28, 97 

\bibitem[\protect\citeauthoryear{Hu et al.}{1994}]{hu94} Hu J.~Y., te Lintel Hekkert P., Slijkhuis F., Baas F., Sahai R., Wood P.~R., 1994, A\&AS, 103, 301 


\bibitem[\protect\citeauthoryear{Imai}{2007}]{ima07} Imai, H.\ 2007, 
in Chapman J.~M., Baan W.~A., eds, Proc. IAU Symp. 242,
Astrophysical Masers \& Their Environments. Cambridge Univ. Press,
Cambridge, p. 279

\bibitem[\protect\citeauthoryear{Imai et al.}{2002}]{ima02} Imai H., Obara K., Diamond P.~J., Omodaka T., Sasao T., 2002, Natur, 417, 829 


\bibitem[\protect\citeauthoryear{Imai et al.}{2004}]{ima04} Imai H., Morris M., Sahai R., Hachisuka K., Azzollini F.~J.~R., 2004, A\&A, 420, 265 

\bibitem[\protect\citeauthoryear{Imai et al.}{2013}]{ima13} Imai H., Deguchi S., Nakashima J.-i., Kwok S., Diamond P.~J., 2013, ApJ, 773, 182 


\bibitem[\protect\citeauthoryear{Kim et al.}{2010}]{kim10} Kim J., Cho S.-H., Oh C.~S., Byun D.-Y., 2010, ApJS, 188, 209 

\bibitem[\protect\citeauthoryear{Lewis, Eder, \& Terzian}{Lewis et al.}{1987}]{lew87} Lewis B.~M., Eder J., Terzian Y., 1987, AJ, 94, 1025 

\bibitem[\protect\citeauthoryear{Lumsden et al.}{2002}]{lum02} Lumsden S.~L., Hoare M.~G., Oudmaijer R.~D., Richards D., 2002, MNRAS, 336, 621 

\bibitem[\protect\citeauthoryear{Meixner et al.}{1999}]{mei99} Meixner M., et al., 1999, ApJS, 122, 221 


\bibitem[\protect\citeauthoryear{Nyman, Johansson, \& Booth}{1986}]{nym86} Nyman L.-A., Johansson L.~E.~B., Booth R.~S., 1986, A\&A, 160, 352 


\bibitem[\protect\citeauthoryear{P{\'e}rez-S{\'a}nchez, Vlemmings, \& Chapman}{2011}]{per11} P{\'e}rez-S{\'a}nchez A.~F., Vlemmings W.~H.~T., Chapman J.~M., 2011, MNRAS, 418, 1402 

\bibitem[\protect\citeauthoryear{Purser et al.}{2016}]{pur16} Purser S.~J.~D., et al., 2016, MNRAS, 460, 1039 

\bibitem[\protect\citeauthoryear{Ramos-Larios et al.}{2009}]{rl09} Ramos-Larios G., Guerrero M.~A., Su{\'a}rez O., Miranda L.~F., G{\'o}mez J.~F., 2009, A\&A, 501, 1207 

\bibitem[\protect\citeauthoryear{Ramos-Larios et al.}{2012}]{rl12} Ramos-Larios G., Guerrero M.~A., Su{\'a}rez O., Miranda L.~F., G{\'o}mez J.~F., 2012, A\&A, 545, A20 

\bibitem[\protect\citeauthoryear{Sahai et al.}{1999}]{sah99} Sahai R., te Lintel Hekkert P., Morris M., Zijlstra A., Likkel L., 1999, ApJ, 514, L115 


\bibitem[\protect\citeauthoryear{Sevenster et al.}{1997}]{sev97} Sevenster M.~N., Chapman J.~M., Habing H.~J., Killeen N.~E.~B., Lindqvist M., 1997, A\&AS, 124,  509

\bibitem[\protect\citeauthoryear{Sevenster et al.}{2001}]{sev01} Sevenster M.~N., van Langevelde H.~J., Moody R.~A., Chapman J.~M., Habing H.~J., Killeen N.~E.~B., 2001, A\&A, 366, 481 

\bibitem[\protect\citeauthoryear{Skrutskie et al.}{2006}]{skr06} Skrutskie M.~F., et al., 2006, AJ, 131, 1163 


\bibitem[\protect\citeauthoryear{Soker}{2014}]{sok14} Soker N., 2014, 
in Morisset C., Delgado-Inglada G., Torres-Peimbert S.,
eds, Proc. Asymmetrical Planetary Nebulae VI Conf., online at http://www.astroscu.unam.mx/apn6/PROCEEDINGS/, 93

\bibitem[\protect\citeauthoryear{Soker}{2016}]{sok16} Soker N., 2016, MNRAS, 455, 1584 

\bibitem[\protect\citeauthoryear{Su{\'a}rez et al.}{2006}]{sua06} Su{\'a}rez O., Garc{\'{\i}}a-Lario P., Manchado A., Manteiga M., Ulla A., Pottasch S.~R., 2006, A\&A, 458, 173 

\bibitem[\protect\citeauthoryear{Su{\'a}rez, G{\'o}mez, \& Miranda}{Su{\'a}rez et al.}{2008}]{sua08} Su{\'a}rez O., G{\'o}mez J.~F., Miranda L.~F., 2008, ApJ, 689, 430

%


\bibitem[\protect\citeauthoryear{te Lintel Hekkert et al.}{1989}]{tel89} te Lintel Hekkert P., Versteege-Hensel H.~A., Habing H.~J., Wiertz M., 1989, A\&AS, 78, 399 

\bibitem[\protect\citeauthoryear{te Lintel Hekkert et al.}{1991}]{tel91} te Lintel Hekkert P., Caswell J.~L., Habing H.~J., Haynes R.~F., Haynes R.~F., Norris R.~P., 1991, A\&AS, 90, 327 

\bibitem[\protect\citeauthoryear{Urquhart et al.}{2009}]{urq09} Urquhart J.~S., et al., 2009, A\&A, 501, 539 

\bibitem[\protect\citeauthoryear{Van de Steene, van Hoof, \& Wood}{Van de Steene et al.}{2000}]{van00} Van de Steene G.~C., van Hoof P.~A.~M., Wood P.~R., 2000, A\&A, 362, 984 


\bibitem[\protect\citeauthoryear{Vlemmings et al.}{2014}]{vle14} Vlemmings W.~H.~T., Amiri N., van Langevelde H.~J., Tafoya D., 2014, A\&A, 569, A92
 
\bibitem[\protect\citeauthoryear{Walker, Matsakis, \& Garcia-Barreto}{1982}]{wal82} Walker R.~C., Matsakis D.~N., Garcia-Barreto J.~A., 1982, ApJ, 255, 128 


\bibitem[\protect\citeauthoryear{Walsh et al.}{1998}]{wal98} Walsh A.~J., Burton M.~G., Hyland A.~R., Robinson G., 1998, MNRAS, 301, 640 


\bibitem[\protect\citeauthoryear{Walsh et al.}{2001}]{wal01} Walsh A.~J., Bertoldi F., Burton M.~G., Nikola T., 2001, MNRAS, 326, 36 

\bibitem[\protect\citeauthoryear{Walsh et al.}{2009}]{wal09} Walsh A.~J., Breen S.~L., Bains I., Vlemmings W.~H.~T., 2009, MNRAS, 394, L70 

\bibitem[\protect\citeauthoryear{Wilson et al.}{2011}]{wil11} Wilson W.~E., et al., 2011, MNRAS, 416, 832 

\bibitem[\protect\citeauthoryear{Wolak, Szymczak, \& G{\'e}rard}{2012}]{wol12} Wolak P., Szymczak M., G{\'e}rard E., 2012, A\&A, 537, A5 

\bibitem[\protect\citeauthoryear{Wright et al.}{2010}]{wri10} Wright E.~L., et al., 2010, AJ, 140, 1868-1881 

\bibitem[\protect\citeauthoryear{Yung et al.}{2011}]{yun11} Yung B.~H.~K., Nakashima J.-i., Imai H., Deguchi S., Diamond P.~J., Kwok S., 2011, ApJ, 741, 94 

\bibitem[\protect\citeauthoryear{Yung et al.}{2013}]{yun13} Yung B.~H.~K., Nakashima J.-i., Imai H., Deguchi S., Henkel C., Kwok S., 2013, ApJ, 769, 20 

\bibitem[\protect\citeauthoryear{Yung, Nakashima, \& Henkel}{Yung et al.}{2014}]{yun14} Yung B.~H.~K., Nakashima J.-i., Henkel C., 2014, ApJ, 794, 81 

\bibitem[\protect\citeauthoryear{Yung et al.}{2017}]{yun17} Yung B.~H.~K., Nakashima J.-i., Hsia C.-H., Imai H., 2017, MNRAS, 465, 4482 

\end{thebibliography}




\appendix

\section{Other OH maser sources detected in the observed fields}

\label{sec:app}
%

In addition to our target sources, we detected OH maser emission from other sources in the field containing IRAS 15544-5332. No OH source was detected in the other three fields. Since these sources are { displaced} from the phase center of the interferometer, the data presented in this appendix have been corrected by the primary beam response.

\subsection{Emission at 1612 MHz}

\begin{figure}
	\includegraphics[width=0.9\columnwidth]{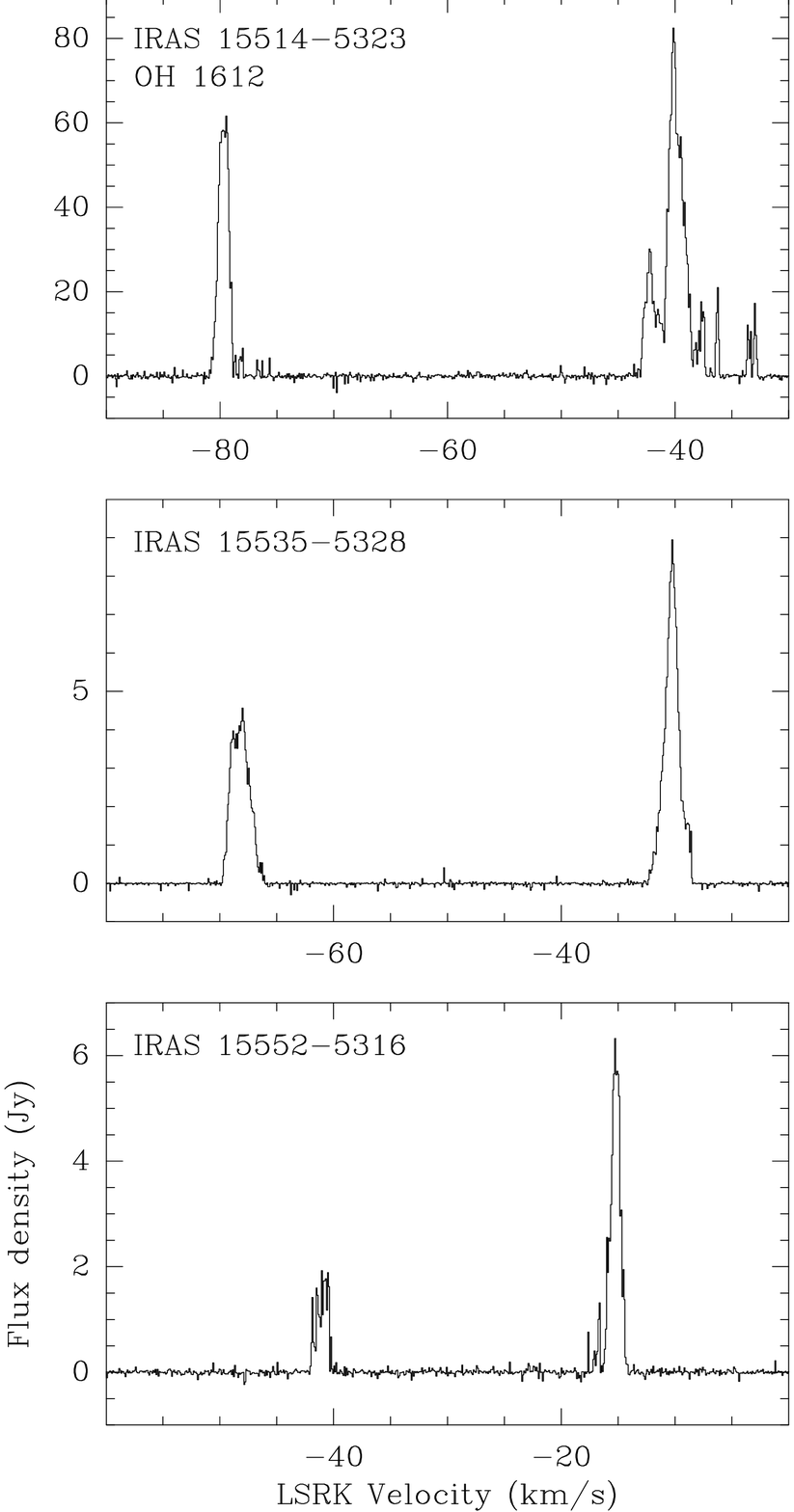}
	\caption{OH maser spectra at 1612 MHz of sources within the IRAS 15544-5332 field. Spectra have been corrected by the primary beam response.}
	\label{fig:spec_i15oh1612}
\end{figure}

OH emission at 1612 MHz was detected from three distinct locations. The most likely sources of excitation, together with the parameters of the emission are given in Table \ref{tab:oh1612other}. The spectra are given in Fig. \ref{fig:spec_i15oh1612}. This OH emission seems to be associated with evolved objects. The spectra roughly show typical double peaked profiles of AGB stars, with velocity separations $\simeq 25-40$ km s$^{-1}$. In the case of  IRAS 15514-5323, there are several additional weaker features near both peaks, more notably around the redshifted one, which could trace mass-loss processes different from the mere expansion of the circumstellar envelope.

\begin{table*}
	\centering
	\caption{Parameters of additional OH maser emission at 1612 MHz in the observed fields.}
	\label{tab:oh1612other}
	\begin{tabular}{lllllllll} %
		\hline
Source &  $S_\nu$	& $V_{\rm LSRK}$	& R.A.(J2000)	& Dec(J2000)	& Error & $\int S_\nu dV$ & $V_{\rm min}$ & $V_{\rm max}$ \\
		&  (Jy) 		& (km s$^{-1}$) & 				&				& (mas) & (Jy km s$^{-1}$) & (km s$^{-1}$) & (km s$^{-1}$) \\
		\hline
IRAS 15514-5323 &  $76.7\pm 0.7$  & $-40.1$	& 15:55:20.24 & $-53$:32:40.7 & 500 & $237.6\pm 1.5$ & $-80.9$ & $-32.7$\\
IRAS 15535-5328	&  $8.76\pm 0.04$ & $-30.2$	& 15:57:23.431 & $-53$:37:30.01 & 160 & $21.19\pm 0.09$ & $-69.8$ & $-28.4$\\ 
IRAS 15552-5316 &  $6.39\pm 0.06$ & $-15.3$	& 15:59:07.096  & $-53$:25:03.73 & 280 & $8.83\pm 0.09$ & $-42.1$ & $-19.4$ \\ 
		\hline
	\end{tabular}
\end{table*}

OH emission at 1612 MHz from IRAS 15514-5323 (OH 328.225+0.042), IRAS 15535-5328 (OH 328.4-0.2), and IRAS 15552-5316 (OH 328.7-0.2)
has been reported before \citep[e.g.,][]{cas75,gay88,tel91,sev97}, although our data have a higher spectral resolution. 

\subsection{Emission at 1665 and 1667 MHz}

\begin{figure}
	\includegraphics[width=0.9\columnwidth]{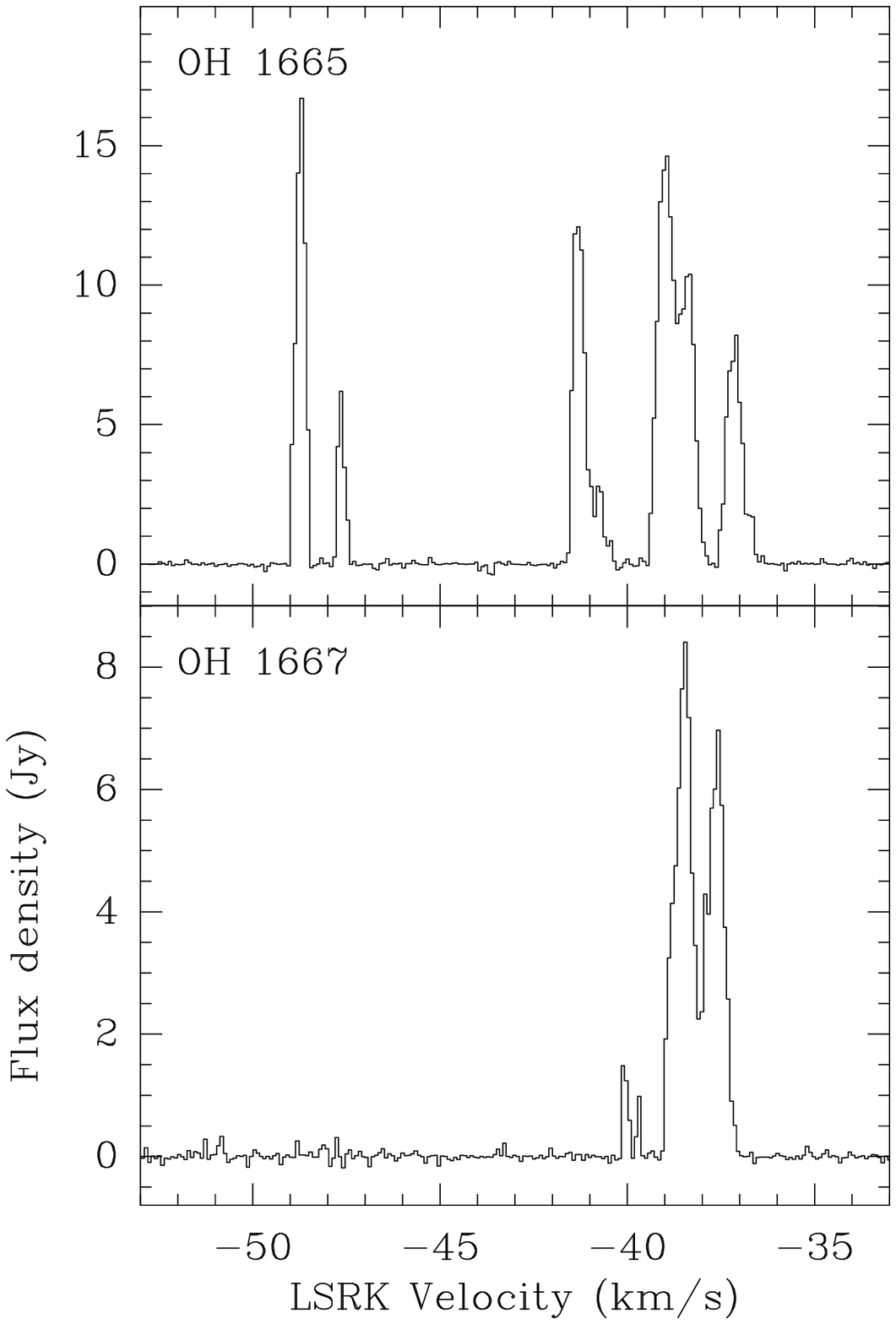}
	\caption{OH maser spectrum at 1665 and 1667 MHz of sources within the IRAS 15544-5332 field. { The spectra show the combined emission from sources Caswell OH 328.237-0.547 and Caswell OH 328.254-0.532, which are both within the synthesized beam of the images.} Spectra have been corrected by the primary beam response.}
	\label{fig:spec_i15oh1665}
\end{figure}

The emission at these frequencies seems to arise from { several individual} sources. { However,} their relative distances are smaller than the size of the synthesized beam in our images, so it is impossible to obtain a spectrum for each of the individual sources. However, the relative positional accuracy of each individual spectral feature is enough to discriminate their positions. Fig. \ref{fig:spec_i15oh1665} shows the spectra at 1665 and 1667 MHz. { Both spectra represent the combined emission from the different sources within the synthesized beam.}

The emission at these frequencies was better resolved by \cite{cas98} using a more extended ATCA configuration. It corresponds to sources 328.237-0.547 and 328.254-0.532 in that paper, and it could be associated with mid-infrared sources 49, 50, and 51 in \cite{wal01}. These are sites of methanol maser emission \citep{wal98} and therefore, they are likely to be high-mass star-forming regions.

\bsp	
\label{lastpage}
\end{document}